\documentclass[aps,prb,preprint,superscriptaddress]{revtex4-1}
\usepackage{graphicx}
\usepackage{bm}
\usepackage{hyperref}
\usepackage{amsmath}
\usepackage{times}
\usepackage{graphicx}
\usepackage{multirow}
\begin{document}

\title{Ehrlich-Schw\"obel Effect on the Growth Dynamics of GaAs(111)A surfaces}

\author{Luca Esposito}
%\affiliation{L--NESS and Dept. of Materials Science, University of Milano-Bicocca, Via R. Cozzi 55, I*--20125 Milano, Italy}
\affiliation{L--NESS and Dept. of Materials Science, University of Milano-Bicocca, Milano, Italy}

\author{Sergio Bietti}
\affiliation{L--NESS and Dept. of Materials Science, University of Milano-Bicocca, Milano, Italy}

\author{Alexey Fedorov}
\affiliation{L--NESS and CNR-IFN Institute, Como, Italy}

\author{Richard N\"otzel}
\affiliation{South China University of Technology, Guangzhou, China}
\affiliation{L--NESS and Dept. of Materials Science, University of Milano-Bicocca, Milano, Italy}

\author{Stefano Sanguinetti}
\email{stefano.sanguinetti@unimib.it}
\affiliation{L--NESS and Dept. of Materials Science, University of Milano-Bicocca, Milano, Italy}
\affiliation{L--NESS and CNR-IFN Institute, Como, Italy}

\date{\today}

\begin{abstract}
We present a detailed characterization of the growth dynamics of Ga(Al)As(111)A surfaces. We develop a theoretical growth model that well describes the observed behavior on the growth parameters and underlines the Ehrlich-Schw\"obel barrier as leading factor that determines the growth dynamics. On such basis we analyze the factors that lead to the huge observed roughness on such surface orientations and we identify the growth conditions that drive the typical three-dimensional growth of Ga(Al)As(111)A towards atomically flat surface. GaAs/Al$_{0.30}$Ga$_{0.70}$As
quantum wells realized on optimized surface ($<0.2$ nm roughness) show a record low emission
linewidth of 4.5 meV.
\end{abstract}
\maketitle

\section{Introduction}

In the field of semiconductor nanotechnology for transport and optoelectronic
applications there are cases where the employment of non-(100) surfaces
improves the characteristics of the device or can even be necessary
condition for its functioning. Amongst crystalline surfaces the (111)
it's the one that has gathered more attention in the last years, especially
for what concerned the opto-electronic applications of III-V semiconductors.
Its physical and geometrical characteristics, in fact, are particularly
suitable for devices as high mobility transistors (HMT) \cite{Miyata2011,Xu2009,Hudait2013}
and quantum well (QW) intersubband photodetectors \cite{Li2005}, but
also for the feasible implementation of a new generation of devices
like the ones based on topological insulators \cite{Zhang2013a}, spintronics \cite{HBH12,Balocchi2011,Weng2013,Ye2012,Zhao2014}
and entangled photons \cite{Mano2010,Stock2010,Huo2013b,Karlsson2010,Dupertuis2011,Kuroda2013,Singh2010}

However, although high control on surface growth phenomena is a fundamental
factor to avoid spurious and detrimental effects, like fine structure
splitting for entangled photon generation or carriers mobility reduction
for HMT, studies regarding the growth of this surface are few, lacking
of an in depth description of growth mechanisms. 

The (111) surface morphology, in most of the
cases, is still affected by a large surface roughness.
In particular, the growth of highly flat GaAs and AlGaAs (111)A surfaces, and consequently
of high quality nanostructures grown on this substrate, is strongly
subordinated to the possibility of inhibiting the formation of large
(with $\mu$m lateral dimensions) pyramidal hillocks with threefold
symmetry, nucleated by stacking faults \cite{Uehara2007,Horikoshi2007}. 

As a matter of fact, the growth, via Molecular Beam Epitaxy (MBE), of Al$_{x}$Ga$_{1-x}$As (111)A surfaces, with $0\le x \le 0.3$, shows a complex behaviour. A full
explanation of the phenomenology cannot be given without a careful
study of islands growth dynamics.  
Here present the systematic study of the growth of Ga(Al)As(111)A to identify
prominent adatom incorporation mechanisms, the model we built to understand
and control the growth along this crystallographic direction, and
the growth procedure we used to obtain atomically flat ($<$0.2
nm RMS)  Al$_{0.30}$Ga$_{0.70}$As and GaAs (111)A surfaces. 
A fundamental role in determining the morphology of GaAs(111)A surfaces is played by the presence of a sizeable Ehrlich-Schw\"obel barrier\cite{Johnson1994,MicKru04} which promotes the three--dimensional growth. The understanding of the growth dynamics on Ga(Al)As (111)A surfaces allowed us to obtain a quantum well with extremely narrow emission linewidth on substrates with [111] orientation. 

\section{Methods}

We fabricated and morphologically characterized twelve samples divided in three series, one for each of the free growth parameters we explored: i) substrate temperature (T), ii) group-III flux ($F$) and iii) V/III
BEP ratio ($\Phi /F$). See Table \ref{tab:Growth} for a summary of the sample
growth conditions. 
All samples were grown by a molecular beam epitaxy (MBE) system on
a (111)A semi-insulating GaAs substrate, with a nominal structure
composed by a 10nm GaAs buffer layer and a 100nm Al$_{0.30}$Ga$_{0.70}$As
epilayer. Both the buffer and the Al$_{0.30}$Ga$_{0.70}$As epilayer were grown at the same temperature. 

After the growth, each sample was analyzed by an atomic force microscope
(AFM) working in tapping mode, using tips with radius of curvature of 2 nm for the morphological characterization.  The photoluminescence measurements of the fabricated QW was conducted at the temperature of 15 K 
and excited by Nd:YAG laser, doubled in frequency, at 532 nm with spot diameter of $\sim$300 $\mu$ m.

\begin{table}
\begin{tabular}{|c||c|c|c|c|}
\hline 
{\Large{}\textbackslash{}} & sample & T($^\circ$C) & group-III flux (s$^{-1}$ cm$^{-2}$) & V/III ratio\tabularnewline
\hline 
\hline 
\multirow{4}{*}{
Temperature
} & T1 & 460  & 6$\cdot10{}^{14}$ & 75\tabularnewline
\cline{2-5} 
 & T2 & 490  & 6$\cdot10{}^{14}$ & 75\tabularnewline
\cline{2-5} 
 & T3 & 520  & 6$\cdot10{}^{14}$ & 75\tabularnewline
\cline{2-5} 
 & T4 & 550  & 6$\cdot10{}^{14}$ & 75\tabularnewline
\hline 
\multirow{4}{*}{
Growth Rate} & G1 & 520  & 0.5$\cdot10{}^{14}$ & 75\tabularnewline
\cline{2-5} 
 & G2 & 520  & 1.5$\cdot10{}^{14}$ & 75\tabularnewline
\cline{2-5} 
 & G3 & 520  & 3$\cdot10{}^{14}$ & 75\tabularnewline
\cline{2-5} 
 & G4 & 520  & 6$\cdot10{}^{14}$ & 75\tabularnewline
\hline 
\multirow{4}{*}{
V/III Ratio
} & R1 & 520  & 6$\cdot10{}^{14}$ & 75\tabularnewline
\cline{2-5} 
 & R2 & 520  & 3$\cdot10{}^{14}$ & 150\tabularnewline
\cline{2-5} 
 & R3 & 520  & 1.5$\cdot10{}^{14}$ & 300\tabularnewline
\cline{2-5} 
 & R4 & 520  & 0.5$\cdot10{}^{14}$ & 900\tabularnewline
\hline 
\end{tabular}

\caption{Conditions used for the growth of the samples used in this work. Temperature, Growth Gate and V/III BEP Ratio series are indicated.}
\label{tab:Growth}
\end{table}

\section{Results}
\label{results}

\begin{figure}[htbp]
	\centering
		\includegraphics[width=0.50\textwidth]{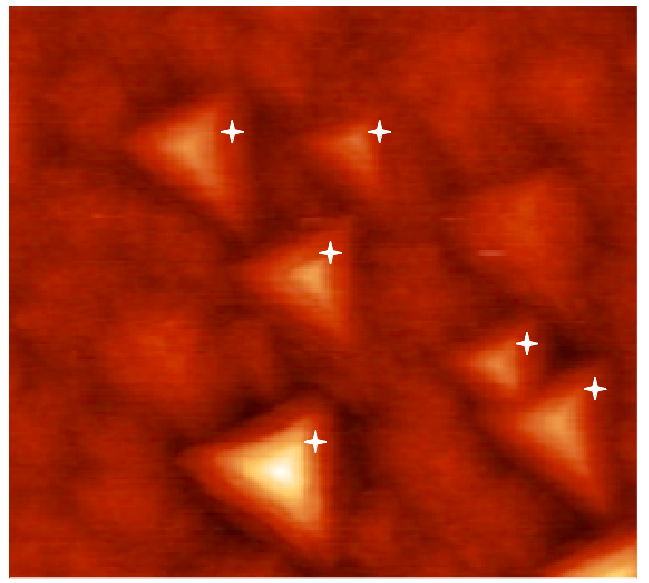}
	\caption{AFM topographical image ($3 \times 3$ $\mu$m$^2$) of sample G2. Hillocks are indicated by white stars.}
	\label{fig:hillock}
\end{figure}

A typical AFM topographical image is presented in Fig. \ref{fig:hillock}.
We can clearly identify two families of pyramidal shaped structures
with different size and aspect ratio (the ratio between vertical over
lateral dimensions), which are present in all the grown samples. 

We will refer to the smaller structures as ``islands'' and to the
big ones as ``hillocks''. The roughness (in terms of RMS) due to
islands is  quite small ($\sim$0.2$^{-1}$ nm on 8x8
$\mu$m$^{2}$ area) while the contribution given by hillocks can
be huge ($\sim$2 nm) and their minimization is one of the points on
which our study is focused on.

It's worth noting that all the results we obtained with Al$_{0.30}$Ga$_{0.70}$As
epilayer are also valid for Al content lower than the $30\%$. 

\subsection{Temperature Series}

\begin{figure}
	\includegraphics[scale=0.80]{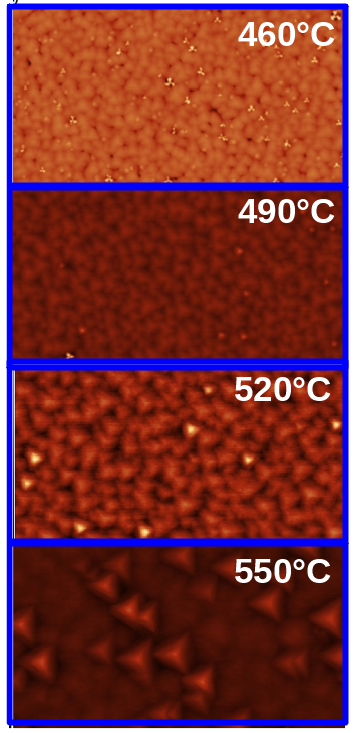}
	\caption{AFM scans (4 $\times$ 8 $\mu$m$^2$) of the samples belonging to the temperature series: T1 (a), T2 (b), T3 (c) and T4 (d). The height scale in the images is from dark to bright with a maximum value of: 30 nm (a), 21 nm (b), 7 nm (c) and 10 nm (d). Growth parameters are reported in Table \protect\ref{tab:Growth}. The temperature ($^\circ$C)  used in the growth is indicated in the panel.  }
	\label{fig:TempMezTot1bis}
\end{figure}

We explored the 460-550$^\circ$C temperature range (see Table \ref{tab:Growth}). 
At 460$^\circ$C, as reported in Fig. \ref{fig:TempMezTot1bis}a, the growth
of the Al$_{0.30}$Ga$_{0.70}$As epilayer is 3D like and the nucleated
islands have the shape of rounded pyramids. Their average height is
2.5 nm with a base of 230 nm, and density of $1.8\times 10^{9}$
cm$^{-2}$. The hillocks, instead, have a shape that resembles an irregular
shamrock, that is a typical symptom of dendritic growth on surfaces
with $\bf{C}_{3}$ symmetry\cite{Einax2013}. The hillocks average height
is 10 nm with a base of 130 nm and density is $1.3\times 10^{8}$
cm$^{-2}$. The overall roughness (RMS) on a 8$\times$8 $\mu$m$^{2}$ area
is around 1.17 nm, while the ``background RMS'' due to islands corrugation
is around 1nm.

Raising the temperature up to 490$^\circ$C (Fig. \ref{fig:TempMezTot1bis}b), the sample growth enters
in a different regime. The islands and hillocks are regular
triangular pyramids with net and definite sides. The average height,
base and density of the island are 1.2 nm, 304 nm and $7.8\times 10^{8}$
cm$^{-2}$ respectively; for hillocks we measured 4.4 nm, 225 nm and
$4.5\times 10^{7}$ cm$^{-2}$ respectively. The overall RMS
on a 8$\times$8 $\mu$m$^{2}$ area is decreased to 0.6 nm while the background
RMS is 0.55 nm.

Increasing further the temperature (see Fig. \ref{fig:TempMezTot1bis}c-d) the height and density of islands are reduce but their base increases. A different behavior can be seen for hillocks, for
which height increases as well as the base (although the aspect ratio
still shows a reduction). Hillock density does not change significantly within the temperature series. The overall RMS increases with the temperature due to the larger contribution from the hillocks to the
value of 1.18 nm at 550$^\circ$C. The background RMS due to islands decreases to 0.4 nm.

\subsection{Grow rate series}
\label{FSer} 

\begin{figure}
	\includegraphics[scale=0.80]{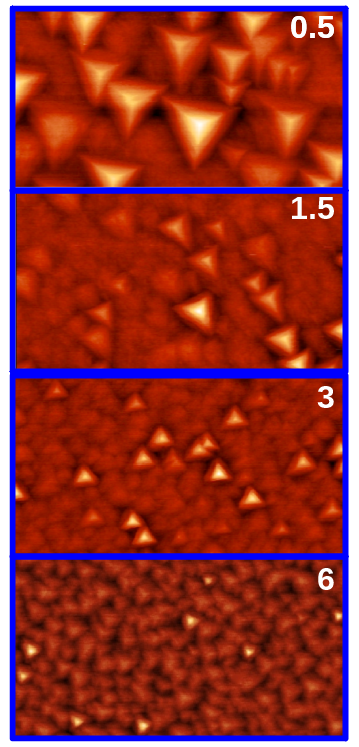}
	\caption{AFM scans (4 $\times$ 8 $\mu$m$^2$) of the samples belonging to the flux series: G1 (a), G2 (b), G3 (c) and G4 (d). The height scale in the images is from dark to bright with a maximum value of: 16 nm (a), 12 nm (b), 11 nm (c) and 7 nm (d). Growth parameters are reported in Table \protect\ref{tab:Growth}. The group III flux (10$^{14}$ cm$^{-2}$s$^{-1}$) used in the growth is indicated in the panel.}
	\label{fig:FluMezTot1bis}
\end{figure}

We explored the $0.5-6\times 10^{14}$ atoms cm$^{-2}$
s$^{-1}$growth rate range (referred to group III adatoms) by changing
simultaneously group III and group V flux but maintaining a constant V/III ratio (see Table \ref{tab:Growth}). At a rate of $0.5\times 10^{14}$
atoms cm$^{-2}$ s$^{-1}$ big hillocks are observable with heights
of more than 11 nm and base of 1.5 $\mu$m (Fig. \ref{fig:FluMezTot1bis}a). The areas free from hillocks
are quite small (less than 2$\mu$m$^{2}$) and on the latter the
islands are almost completely merged. Because of that, no reliable
statistics on islands can be made, although a low background roughness
can be inferred from the 0.38 nm background RMS (obtained excluding
hillocks)

Increasing the growth rate (Figs. \ref{fig:FluMezTot1bis}b-d), hillocks become smaller, keeping their
density basically constant. Accordingly, the overall RMS decreases
till the value of 0.57 nm at $6 \times 10^{14}$ atoms cm$^{-2}$
s$^{-1}$. For what concerns islands, the increase of the growth rate
results in larger density and aspect ratio. This leads to
a rise of the background RMS.

\subsection{V/III ratio series}

\label{RSer} 

\begin{figure}
	\includegraphics[scale=0.80]{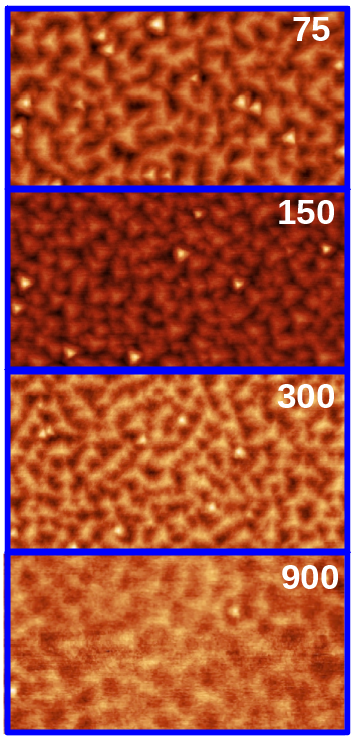}
	\caption{AFM scans (4 $\times$ 8 $\mu$m$^2$) of the samples belonging to the flux series: R1 (a), R2 (b), R3 (c) and R4 (d). The height scale in the images is from dark to bright with a maximum value of: 3 nm (a), 7 nm (b), 5 nm (c) and 2.5 nm (d). Growth parameters are reported in Table \protect\ref{tab:Growth}. $\Phi/F$, measured as As to Ga BEP ratio, used in the growth is indicated in the panel.}
	\label{fig:RatMezTot1bis}
\end{figure}

The series spans 75-900 BEP V/III ratio range. At a ratio
of 75, that is the one used for to realize the Temperature and Grow Rate series, the
growth is 3D like with clearly distinguishable pyramidal hillocks.
The overall RMS is around 0.57 nm and the background one is around
0.47 nm (Fig. \ref{fig:RatMezTot1bis}a).

Decreasing the group III flux, in order to increase the V/III ratio
till 300 (Fig.\ref{fig:RatMezTot1bis}c), only a small change is observed in island density, which
only reduces by a factor two, from 6 to $3.6\times 10^{8}$ cm$^{-2}$,
when the V/III ratio increases from 75 to 300. Hillocks density is
around $3\times 10^{7}$ cm$^{-2}$ in the whole range. However,
the height of the two pyramid families decrease proportionally to
V/III passing from an average value of 1.4 nm to 0.7 nm for islands,
and from a value of 4.5 nm to 2.26 nm for hillocks.

\begin{figure}
	\centering
		\includegraphics[width=0.80\textwidth]{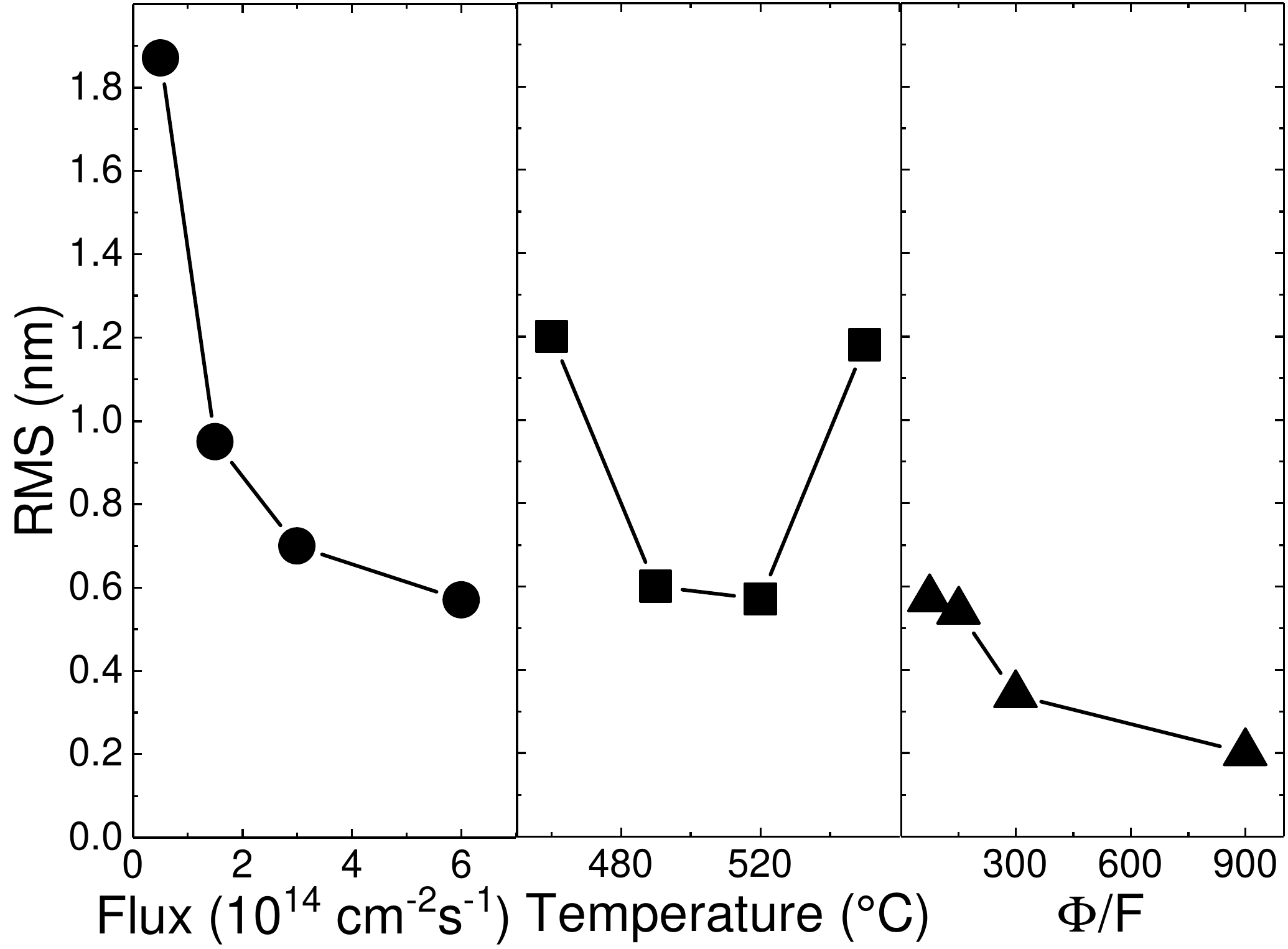}
	\caption{Dependence of overall RMS on the relevant growth parameters in Growth Rate (left panel), temperature (center panel) and V/III Ratio (right panel) series, respectively.}
	\label{fig:rms}
\end{figure}

Going further, to a V/III ratio of 900 (Fig. \ref{fig:RatMezTot1bis}d) the regime of growth passes from
a 3D to a 2D like. The islands are practically one mono-layer high
and the hillocks, if present, are completely indistinguishable from
islands. 
The overall RMS (and obviously the background RMS) drops to a value
of 0.2 nm. The dependence of the RMS on the relevant growth parameters in the three series is reported in Figs. \ref{fig:rms}.

\section{Discussion}

\begin{figure}
	\centering
		\includegraphics[width=0.80\textwidth]{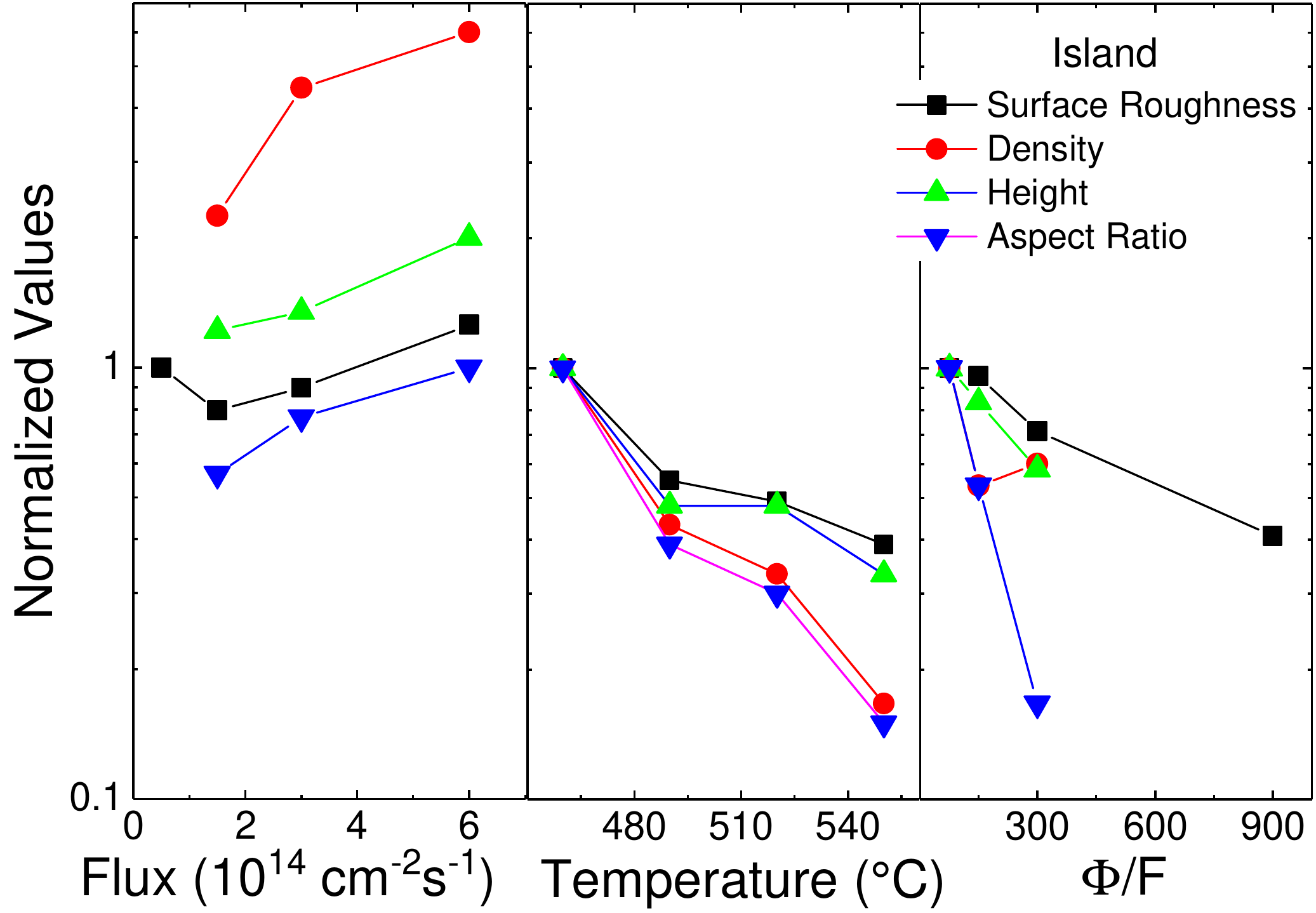}
	\caption{ Normalized dependence on the relevant growth parameter of island RMS (black squares), Density ($\rho$, red circles), height (green up-triangles) and aspect ratio (blue down-triangles) of the series and $F$ (left panel), T (center panel) and $\Phi /F$  (right panel) . The data are normalized to the one assumed at lowest value of the relevant growth parameter in the series}
	\label{fig:trends}
\end{figure}

The description of the observed behavior of surface morphology, in terms of island RMS, density ($\rho$), height and aspect ratio (height over base ratio) is reported in Fig. \ref{fig:trends} .
Some general trends can be identified. All the island relevant parameters tend to decrease with the increasing temperature and group-V ($\Phi$) over group-III ($F$) ratio, while the trend is reversed in the case of $F$ dependence .

As a matter of fact, the variety of morphological observations in the three series resembles
quite well what has been already reported in literature about the homoepitaxial
growth on (111) oriented metals (Pt, Cu, Ag)\cite{Rosenfeld1997} with an fcc
lattice.  In these materials the growth dynamics is quite well understood
and extensive models based on a kinetic growth framework  have been developed \cite{MicKru04}. Within this scenario, it is natural
to use the same basic concepts borrowed from the homoepitaxial growth of metals in the attempt of building an extended
model for binary-alloy, like III-As materials, that could keep the
same ``predictive power'' of the one made for metals. 

The fundamental ingredient of (111) metal homoepitaxial models is the presence of a strong
effective Ehrlich-Schw\"obel (ES) barrier that, frustrating 
the escape of deposited adatoms from the top of a 2D nucleated island, drives the growth towards 3D. We suppose that the same phenomenon is playing a role in the observed behavior on GaAs.  
An estimation of the ES-barrier on Ga(Al)As (111)A is not available in the literature. However, a lower
bound for the ES-barrier will follow directly from \emph{{a--posteriori}} considerations
after the construction of the model. 

Before entering into the construction of the model it is worth to
give a qualitative picture of what happens during the epitaxial
growth, in order to fix some fundamental ideas and define the needed
parameters.

In a layer-by-layer regime, after an ``initial delay time'', the adatoms
impinging on the substrate surface start to aggregate and to form
small 2D islands. After the nucleation of 2D islands, depending on
the surface coverage and on single island size, the atoms coming from
the cells have a certain probability of direct impingement onto the
top of the freshly formed 2D islands.

In this case, after a time $\tau$ called \textit{residence time},
on which the adatom explores the island surface, three events can occur:
1)the adatom re-evaporate; 2) the adatom descends from the island top and eventually sticks on
the step-edge or starts the nucleation of a new island; 3) the adatom
meets other adatoms and starts to nucleate a new island on the top
of the previous one.  Being in a complete condensation regime we will completely discard  re-evaporation processes. Depending on which of the two remaining events dominates,
the growth will proceed 2D or 3D like.

Assuming the case where the adatom diffusion is quite efficient and
therefore the diffusion length $l_{D}$ is larger than the average equivalent
radius $R$ of the 2D islands ($l_{D}\gg R$), the factor that can
lead to multilayer nucleation is the ES-barrier. A large value of the ES-barrier,
in fact, induces an increase of the residence time $\tau$ and an increase
of the probability of adatoms encountering on the top of the island. 

Within this kinetic framework three classes of parameters are important
for the determination of the growth dynamics (see \cite{MicKru04}):
\begin{itemize}
\item the ones related to intralayer diffusion;
\item the ones related to ES-barrier; 
\item the ones related to the growth parameters.
\end{itemize}

In the first class we find the diffusion length $l_{D}$, the diffusion
time $\tau_{D}$ and the diffusivity $\mathrm{D}$. These parameters
are related one to each other by the formula $l_{D}=\sqrt{4\mathrm{D}\tau_{D}}$,
where the rate for a diffusion process, depending on the temperature
$T$ and on the diffusion barrier $E_{D}$, is given by the law $\nu=\nu_{0}e^{-\frac{E_{D}}{kT}}$,
with $\nu_{0}$ is the attempt frequency.

In the second class we find the ES length $l_{ES}=\left(\frac{\nu}{\nu'}-1\right)a$
and the ES time $\tau_{ES}=\frac{R}{2a\nu'}$, where $a$ is the lattice
parameter and, similar to what is seen for the diffusion process, $\nu'$
is the rate of descending from step-edge regulated by the law $\nu'=\nu_{0}'e^{-\frac{E_{S}}{kT}}$,
where $E_{S}$ is the sum of the $E_D$ and of the ES-barrier $\Delta E_S$.

In the growth parameters class we have the temperature $T$, the flux
of Ga $F$ and the flux of As $\Phi$.

\subsection{Description of the model}

It's worth noting that the all growths were conducted
in group III limited conditions (As overpressure). In these conditions,
once the As pressure is effectively taken into account in the
value of diffusivity, the growth rate is completely determined by
the group III (Ga and/or Al) flux $F$ only. For these reasons, using the term ``adatom''
we will refer to group III adatoms only.

Moreover, similarly to what has been done for metals\cite{MicKru04}, we
will assume that in our growth conditions the growth dynamics is not dominated by intrinsic (extremely low diffusivity) or extrinsic
(high growth rate) limitation in the adatoms diffusion path. These
assumptions are not free from consequences and impose important bounds
between the relevant time scales of our model. 

In order to discard a diffusivity limited dynamics,
we should impose that the time elapsed by the adatom on the island edge trying
to overcome the ES barrier and be incorporated in the island edge $\tau_{ES}$ is larger than the time needed to explore
the island top surface $\tau_{D}$: $\tau_{ES}\gg\tau_{D}$.
This way the residence time $\tau$, which is roughly
the sum of the two contributions ($\tau_{D}^i$ and $\tau_{ES}$), satisfies the condition $\tau\simeq\tau_{ES}$, that is diffusivity dynamics plays only a minor role in determining the residence time. This condition remains true till the multilayer nucleation takes
place on the top of a layer (for the second layer is the base of the
island, for the third is the second layer and so on) with radius $R\ll R_{cD}$,
where $R_{cD}$ is the critical radius for which $\tau_{ES}=\tau_{D}$:

\begin{eqnarray}
\frac{R_{cD}^{2}}{4D}=\frac{R_{cD}}{2a\nu'} & \longmapsto & R_{cD}=\frac{2D}{a\nu'}.\label{eq:1}
\end{eqnarray}

To discard dynamics limited by high growth rates, we
should impose that the average delay time $\Delta t=\frac{1}{F\pi R^{2}}$
between the arrival of two adatoms on the top of a 2D island of radius $R$ satisfy
the relation $\tau\ll\Delta t$. This relation also reflects on the
island radius giving the relation $R\ll R_{cG}$ with $R_{cG}$ given
by:

\begin{eqnarray}
\frac{R_{cG}}{2a\nu'}=\frac{1}{F\pi R_{cG}^{2}} & \longmapsto & R_{cG}=\sqrt[3]{\frac{2a\nu'}{F\pi}}.\label{eq:2}
\end{eqnarray}

Borrowing from metals\cite{MicKru04} a value of $\nu'$ on the order
of $10^{6}$Hz, from GaAs (001)\cite{Bietti2014a} a value at $520^\circ$C
for $D$ of $2.24\times10^{7}$cm$^{2}$s$^{-1}$ and setting $F$ = $6\times10^{14}$cm$^{-2}$s$^{-1}$,
$a=3.25\textrm{\AA}$, we get values for $R_{cD}$ and $R_{cG}$ of
$\sim100$ nm and $\sim45$ nm respectively. Although these values
are just indicative, because the used values of $\nu'$ and $D$ for are just educated guesses of what happens on AlGaAs(111)A surfaces, they anyway
show a sort of  internal consistency of our model. In fact, in addition to being both in a "reasonable" range for the phenomena we are considering, they are also of the same order of the average radius that can be infered from the AFM measurements.
The formation of a nucleus on the second layer requires, first, that (at least) two atoms are present on the island simultaneously, and, second, that the two meet before one of them escapes from the island. 

Under the assumption 
$\tau_{D}\ll\tau\ll\Delta t$, the probability
that an adatom, deposited at a time $t=0$ is present on top of a 2D island, at a time $t_{1}$ later
than the arrival time $t_{2}$ of a second atom, can be approximated
as follows \cite{MicKru04}:
\begin{alignat}{1}
P[t_{1}>t_{2}] & =\frac{1}{\Delta t}\int_{0}^{\infty}P_{_{res}}(t_{1})dt_{1}\int_{0}^{t_{1}}e^{-\frac{t_{2}}{\Delta t}}dt_{2}\label{eq:3}\\
 & \approx\frac{1}{\Delta t}\int_{0}^{\infty}t_{1}P_{_{res}}(t_{1})dt_{1}\nonumber \\
 & \approx\frac{\tau}{\Delta t}.\nonumber 
\end{alignat}

In the case of strong step edge barriers the two atoms will almost certainly meet, because their residence time $\tau$ is much larger than the encounter time, which can be roughly identified with the diffusion time $\tau_D$. The probability for a freshly deposited atom to form a nucleus is therefore equal to the probability that another atom is present on the island at its deposition.
Multiplying this probability for the average arrival rate on the island top $\frac{1}{\Delta t}$, we get the 
rate

\begin{equation}
\omega=\frac{\tau}{\Delta t^{2}}=\frac{\pi^{2}F^{2}R^{5}}{2a\nu'}\label{eq:4}
\end{equation}
 
that two adatoms are present at the same time on top of an island. The rate $\omega$ is intuitively proportional to second layer nucleation
rate, except in the special case for which the critical nucleus $i^{*}=1$
(two adatoms for nucleation). In this case in fact $\omega$ will
be exactly equal to second layer nucleation rate and consequently
Eq. \eqref{eq:4} will return the value of the latter. 

This rate changes very fast depending on island radius, passing for
example (using the previously estimated values for $\nu'$ and $D$) from $\sim10^{-3}$ Hz when $R=1$ nm, to a value of $\sim10^{2}$Hz
at $R=10$ nm.

The dependence on growth parameters $T$ and $\Phi$ in Eq. \eqref{eq:4}  is still
hidden, and few more steps are needed to make this dependence explicit. 

The dependence of the average radius $R$ on the growth parameters
can be given by the formula:

\begin{equation}
R=R(F;\Phi;T;\Theta)=\sqrt[3]{\left(\frac{\mu V}{\rho}\right)}=\lambda\rho(F;\Phi;T;\Theta)^{-\frac{1}{3}},\label{eq:5}
\end{equation}

where$\lambda$ is constant, $V$ is the volume of group III elements,  the $\mu$ value depends on the island shape, $\rho$ is the density of islands and $\Theta$ is the coverage. 
In order to evaluate the dependence of $\rho$ on the growth parameter $F$, $\Phi$ and $T$ we extended Ref. \cite{Venables1984} results, by including the effects of As flux via its influence on the Ga diffusion length. This has been done by adding a power dependence to the island density of the form $\Phi^q$, with $q > 0$  as the effect of As flux is to reduce the Ga diffusion \cite{Bietti2014a} and hence to increase the island density. 

\begin{equation}
\rho(F;\Phi;T) \approx \xi (\Theta ) F^{p}\Phi^{q}e^{\frac{E_{n}}{kT}},\label{eq:6}
\end{equation}

where $\xi (\Theta )$ is a weak function of the coverage, $p= i^*/(i^*+\zeta)$ being Ga adatom kinetics in the complete condensation regime \cite{Venables1984} (with $2<\zeta<2.5$ depending on growth dimensionality\cite{Venables1984}) and $E_{n}$ is the nucleation energy. Using this expression in Eq. \eqref{eq:5} the
following relation for the average radius can be deduced

\begin{equation}
R(F;\Phi;T)=\eta(\Theta) F^{-\frac{p}{3}}\Phi^{-\frac{q}{3}}e^{-\frac{E_{n}}{3kT}},\label{eq:7}
\end{equation}

where $\eta(\Theta)$ is a weak function of the coverage  $\Theta$. Inserting Eq. \eqref{eq:7} in Eq. \eqref{eq:4}
and using the definition of $\nu'$ previously given, we can finally
get 

\begin{equation}
\omega(F, \Phi, T) \approx \alpha(\Theta) F^{2-\frac{5}{3}p}\Phi^{-\frac{5}{3}q}e^{\frac{3E_{S}-5E_{n}}{3kT}}
\label{eq:omega_fin}
\end{equation}

where $\alpha(\Theta)$ is, also in this case, a weak function of the coverage. The island second layer nucleation rate depends on two adatom coexistence at the island top, thus $\omega(F; \Phi; T;\Theta)$  gives a direct access to the dependence of surface roughness on the growth parameters.
It is worth noting that being  the  exponent  $p = i^*/(i^*+\zeta)$,  for any value of $i^*<\infty$,  we get that $\frac{5}{3} p<2$, so that  $\omega$ increases proportionally with the increasing of Ga flux $F$. 
The exponent $q$ is always positive, being the effect of As flux to decrease group III adatom diffusion \cite{Horikoshi2007,Uehara2007}

\subsection{Discussion of the results}

A fundamental ingredient for the discussion is the knowledge of the parameters involved in the model, namely the two exponents $p$ and $q$, the activation energy $E_n$, and the ES barrier energy $E_S$. The a--priori knowledge of their values is precluded to us, being the growth dynamics too complex in the case of III-V compounds.  However, by fitting the island density dependence on the growth conditions, we can obtain a direct measure of at least three of the four parameters involved, as $\rho$ explicitly depends on $T$, $F$ and $\Phi$ (see Eq. \eqref{eq:6}). The obtained values are: $E_n = 0.97 \pm 0.1 eV$, $p= 0.37 \pm 0.31$ and $p+q=0.71 \pm 0.16$ from wich we can infer that $q\approx 0.34$. It is worth noting that, despite the large error bar, the $p$ value is in qualitative agreement with the expected exponent for the dependence on the Ga flux: $1 > p > 0.28$.
Inserting the parameters in Eq. (\ref{eq:omega_fin}) it is possible to derive some general rules regarding second layer nucleation in different growth conditions. A  little more algebra must be performed when looking for the dependence on V/III ratio $(\Phi / F)$. In the case that the ratio is varied at constant $\Phi$ we obtain $ \omega \propto \left(\frac{\Phi}{F}\right)^{{\frac{5}{3}p}-2}$ while in the opposite case (constant $F$) $ \omega \propto \left(\frac{\Phi}{F}\right)^{-\frac{5}{3}q}$. 

\begin{enumerate}
\item second layer nucleation increases with the increasing of Ga flux $F$, being the exponent ${2-\frac{5}{3}p}$ always positive in the complete condensation regime, 
so that 3D growth is favored by higher $F$;
\item second layer nucleation decrease with the increasing of As flux $\Phi$, being $q$ a positive quantity,
so that 3D growth is hindered by higher $\Phi$;
\item second layer nucleation is decreasing function of increasing V/III ratio, being the exponents ${{\frac{5}{3}p}-2}$ and ${-\frac{5}{3}q}$ always negative, so that 3D growth is hindered by higher $\frac{\Phi}{F}$;
\item second layer nucleation decrease with the increasing of temperature
$T$, so that 3D growth is hindered by higher $T$, if  $3E_{S}-5E_{n}>0$.
\end{enumerate}
 
Following the considerations made at the end of the last sub-section,
we can interpret the morphological observations as a function of the growth parameters by means of Eq. \eqref{eq:omega_fin}. Concentrating our attention on the islands, we clearly see that our model well describes the experimental results, being the surface roughness an increasing function of $F$ and a decreasing function of $\Phi$ and $\frac{\Phi}{F}$ (rules 1-3).
In the temperature series (see Fig. \ref{fig:TempMezTot1bis}) 3D growth is reduced for the islands by increasing the substrate temperature.  This observation sets the ES energy value $E_S = E_D + \Delta E_S > 5/3 E_n = 1.6$ eV (see rule 4). The value of $E_D$ on GaAs(111)A is not available in the literature. We can have an estimate of it using theoretical calculations on GaAs(001) where $E_D\approx 1.5$ eV. The additional step edge barrier $\Delta E_S = E_S - E_D$ related to the ES edge effect is then expected to be larger than one hundred meV. 

However, when considering the actual overall surface roughness dependence on the growth conditions, we should pay attention to
other factors that cannot be included in the ES-model that 
has brought us to Eq. \eqref{eq:omega_fin}, like extended defects and regime transitions.

A good starting point for such discussion is represented by the temperature series. 
Although the island population shows a reduction of the 3D growth with the increasing $T$, and consequently a lower RMS at high temperatures, this is not what is observed on the surface overall. Hillock growth dynamics follows, in fact, a different behavior.  The
defect nature \cite{Horikoshi2007,Uehara2007} of hillocks tends to promote second layer
nucleation and to increase their capture area proportionally to surface
diffusion. This leads the morphological behavior of hillocks in the
``opposite direction'' with respect to what is expected for islands. However,
this statement is not true for all the temperature range and adopting
the strategy of decreasing the temperature in order to reduce the
adatom diffusivity can be a bad choice. In fact, as previously reported, decreasing the temperature we enter may in a 
growth regime (dendritic regime), where diffusivity is very low, the
step edge reorganization processes are hindered and the kinetic roughening
dominates completely  the final morphology. 

This observation represents a crucial point. It demonstrates, in fact,
that thinking on the transition from a 3D growth to a 2D like only
as a diffusion limited process is too naive and in some way also
incorrect.

\begin{figure}
	\centering
		\includegraphics[width=0.70\textwidth]{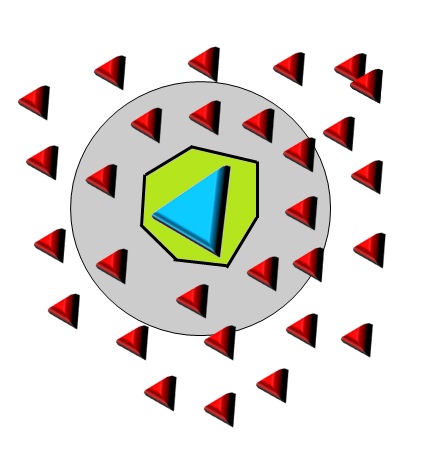}
	\caption{Schematic picture of the reduction of hillocks (large blue island at the center) capture area by adatom mean free control via island (red triangles) density. The effective capture area of the hillock (the green Voronoi area around the hillock) can be strongly reduced, respect to the one determined by the diffusivity (gray circle) by the presence of a high island density.}
	\label{fig:dif-capture}
\end{figure}

For what concerns the growth rate series, we can see again that,
while islands follow the model prediction, that is an increase of the surface roughness with the increasing growth rate, the behavior of hillocks go exactly
in the opposite direction (see Figures \ref{fig:rms} and \ref{fig:trends}). 
The reason is the same discussed for the temperature series, that is a change of the capture area of the hillocks with the growth conditions. Although
in this case the capture area of the hillocks is no more determined by the adatom diffusion. In fact, increasing the growth
rate we are also increasing the island density, due to the increased probability of 
the formation of islands above the critical nucleus size \cite{Venables1984}, thus reducing the inter-island distance well below the diffusion length. This increases the probability that an adatom is incorporated, during its diffusion, into a growing island, thus
decreasing this way the adatom mean free path. This effectively reduces the adatom island capture area to values that can be much lower than the one set by diffusivity only. As shown in Figure \ref{fig:dif-capture}, the effective mean free path is set by the average island-island distance, and this can be controlled by the island nucleation probability and thus, in turn, by the island density. This effect is clearly effecting strongly the hillock growth, as it is determined by the ability of the hillocks to accumulate material from the surroundings. The increase in group III flux, while favouring the second layer nucleation on the islands top, thus increasing island related roughness, it strongly limits the hillocks capture area, thus reducing hillock effects on the overall thickness.

As far as V/III ratio series is concerned, it is worth noting that during the growth of this series the As flux
was kept constant, while the group III was decreased in order to increase
the V/III ratio. We observe that the density of islands decreases when rising
the V/III ratio but, contrary to what is seen in the the previous two
series, the overall RMS decreases too. Looking at Eq. \eqref{eq:omega_fin} we
can see that the island behavior follows what is stated by the model, that is a decrease of second layer nucleation with increasing ($\Phi/F$) ratio. 
What is different from the previous series is that in this case also the hillocks
height decreases with rising the V/III ratio. The reason for this behavior can
be ascribed to the fixed and high As flux. In fact, keeping the As
flux constant at high values we are limiting the diffusivity of group
III adatoms flux and thus, in turn, reducing the hillocks capture zone 
and hindering 3D growth on islands at the same time. Second layer nucleation is, in fact, proportional to $ \omega \propto \left(\frac{\Phi}{F}\right)^{{\frac{5}{3}p}-2}$. The exponent ${{\frac{5}{3}p}-2}$ is in fact always negative, whichever is the critical nucleus size $i^*$.

This is therefore fundamental for the control of the GaAs (111)A surface roughness to find a balance between the reduction of the effective mean free path, obtained both via diffusion length reduction and increase of island density, and the increase of the second layer nucleation, mostly controlled by group III adatom flux. This leads to the necessity to avoid too

\subsection{Quantum Well Optical Quality}

\begin{figure}
	\centering
		\includegraphics[width=0.50\textwidth]{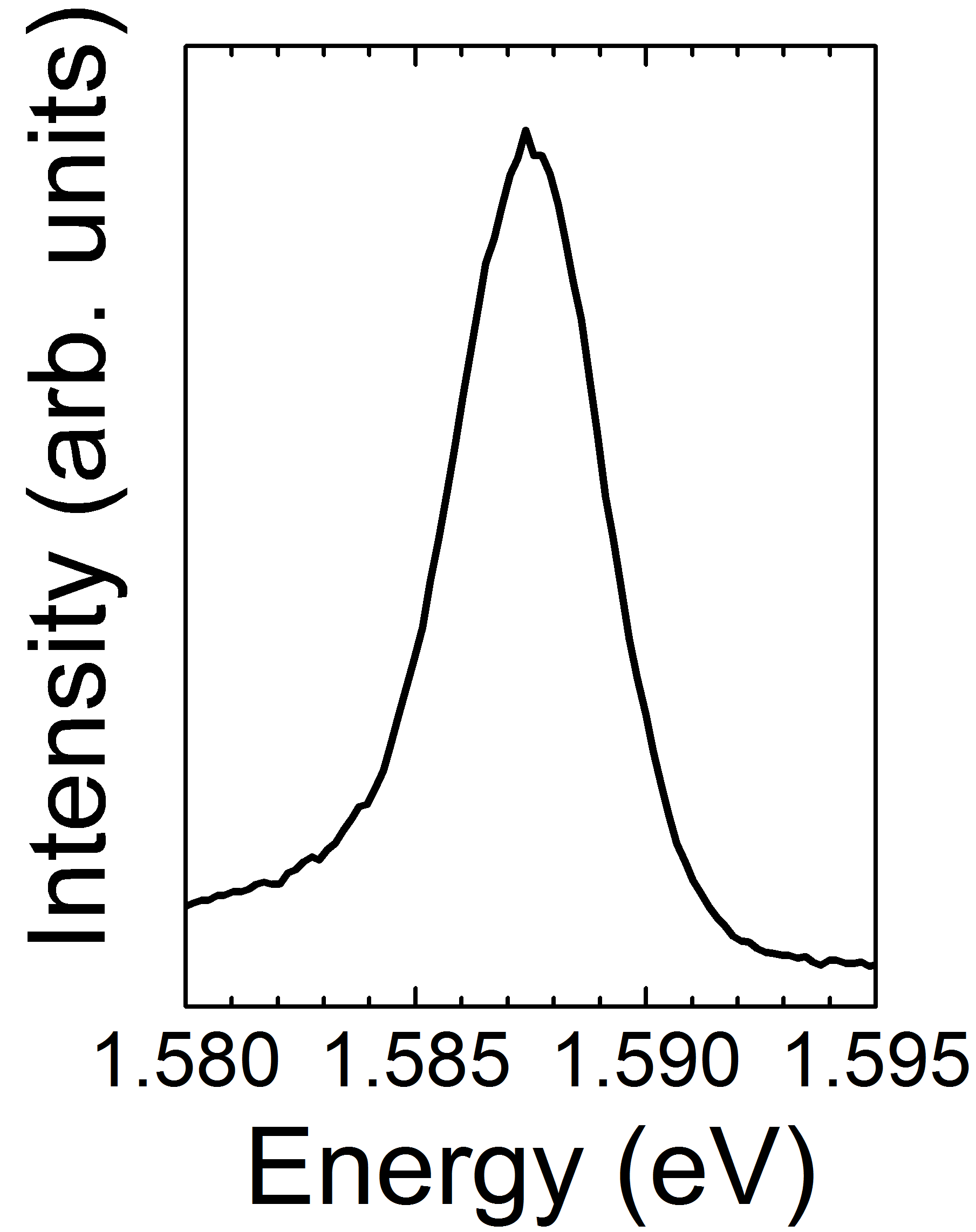}
	\caption{Photoluminescence spectrum of the fabricated 5.5 nm GaAs/AlGaAs QW showing a full width at half maximum of 4.5 meV}
	\label{fig:PL}
\end{figure}

The possibility to obtain extremely flat AlGaAs(111)A surfaces with RMS below $0.2$ nm opens to the fabrication of two dimensional quantum nanostructures with extremely narrow emission. For this purpose we fabricated a 5.5 nm thick, strain free GaA/Al$_{0.3}$Ga$_{0.7}$As quantum well. We expect, in fact, the excitonic linewidth and the AlGaAs/GaAs/AlGaAs interface RMS $dL$ to be related by the equation \cite{Sanguinetti2008a}

\begin{equation}
{dL}\approx \frac{L \cdot dE_{ex}}{2(E_{ex}-E_{g})}
\label{eq-dL}
\end{equation}
where $L$ is the quantum well thickness, $E_{ex}$ is the energy of the interband excitonic transition in the quantum well, $E_g$ the energy gap of the quantum well material and $dE_{ex}$ the emission full width at half maximum. 
The emission spectrum of the fabricated quantum well is shown in Fig. \ref{fig:PL}. The well spectrum show $E_{ex}=1.587$ with $dE_{ex}=4.5$ meV. Using the the experimental values $L=5.5$ $nm$, $E_{ex}=1.587$
eV and $E_{g}=1.515$ $eV$ in Eq. (\ref{eq-dL}) we see that an emission
broadening $dE_{ex}=4.2$ meV corresponds to a well width fluctuation
of $dL=0.17$ nm, in perfect agreement with the roughness value 
$<0.2$ nm measured by AFM.

This is clear demonstration of how the high surface flatness
control that can be reached by growth strategy and the procedures here presented strongly impact on the electronic properties of the quantum nanostructures realized on it. The quantum well in fact shows an extremely sharp emission of only 4.5 meV full width half maximum \cite{cho2003}.

\section{Conclusions}

We presented a detailed study on how MBE growth parameters, namely temperature, growth rate
and V/III ratio, impact on the growth dynamics of Al$_{x}$Ga$_{1-x}$As
($0<x<0.30$) (111)A surface. In particular we have identified and analyzed the factors
that lead to huge overall roughness on this surface: i) the presence
of defect induced hillocks; ii) the island 3D growth. The latter has
been often overlooked in previous studies, thus preventing the achievement
of a truly atomically flat surface \cite{Horikoshi2007,Uehara2007}. 

Our study has identified the Ehrlich-Schw\"oebel barrier as the leading factor on surface adatom dynamics. We then developed a novel theoretical model for the growth of III-V semiconductors in the presence of ES barriers.  Within this framework we were able to interpret and control the island roughness. The hillocks related roughness is controlled by shrinking their capture area  through an efficient mechanism of reduction of adatom mean free path via island density increase. This way we were able to drive the typical 3D growth of Ga(Al)As(111)A towards atomically flat surfaces ($<0.2$ nm RMS). 
Such low surface RMS 
was obtained at a substrate temperature of 520$^\circ$C,
group III flux of 0.5 $10^{14}$ atoms s$^{-1}$ cm$^{-2}$ and V/III
ratio of 900. These growth conditions are quite far from those typical
of (001) surfaces and are dictated by the necessity of suppressing
hillocks growth and promoting 2D growth of islands in the presence of
a large Ehrlich-Schw\"obel barrier. GaAs/Al$_{0.30}$Ga$_{0.70}$As
quantum wells realized on such optimized surface show a record narrow emission
linewidth of 4.5 meV.

The presented modeling of the growth dynamics on Ga(Al)As(111)A surfaces will permit to obtain quantum nanostructures with etremely sharp interfaces, thus reducing excitonic emission broading in QWs and carrier scattering at the interface in HMT fabricated on (111)A substrates, thus opening to the possibility of fabricating high efficiency HMT and optoelectronic devices.

\newpage

\newpage
\begin{center}
\textbf{References}
\end{center}

\printfigures
\newpage
\printtables
\end{document}